# Critical Business Decision Making for Technology Startups: A PerceptIn Case Study

Shaoshan Liu
PerceptIn
shaoshan.liu@perceptin.io

**Abstract**: *most business decisions are made with analysis, but some are judgment calls not susceptible to analysis due to time or information constraints. In this article, we present a real-life case study of critical business decision making of PerceptIn, an autonomous driving technology startup: in early years of PerceptIn, PerceptIn had to make a decision on the design of computing systems for its autonomous vehicle products. By providing details on PerceptIn's decision process and the results of the decision, we hope to provide some insights that can be beneficial to entrepreneurs and engineering managers in technology startups.*

**Background**: PerceptIn was established in 2016 to develop visual perception technologies for autonomous vehicles and robots.  Since its inception, PerceptIn has successfully attracted over 10 million USD of venture capital funding, from Walden International, Matrix Partners, and Samsung Ventures [1].  PerceptIn is an international technology startup with operations in the U.S., Japan, Europe, and Asia. PerceptIn consists of over 30 researchers and engineers and 10 business professionals.  The business professionals are responsible for business development in different markets and gather feedbacks for the company's R&D efforts, whereas the engineers and researchers are responsible for developing cutting edge autonomous driving technologies. In the past three years, PerceptIn has generated over 20 U.S. patents and over 100 international patents, as well as numerous research papers.

In 2017, PerceptIn decided to develop low speed autonomous vehicles to serve the micromobility market, as micromobility is a rising transport mode wherein lightweight vehicles cover short trips that massive transit ignore [2]. According to US Department of Transportation, 60% of vehicle traffic is attributed to trips under 5 miles [3]. Transportation needs in short trips are disproportionally under-served by current mass transit systems due to high cost, which affects the society profoundly. Micromobility bridges transit services and communities' needs, driving the rise of Mobility-as-a-Service.

**Business Analysis**: PerceptIn's primary customers are autonomous vehicle operators around the globe, and PerceptIn partners with these autonomous vehicle operators to provide micromobility services in different markets, such as Japan, U.S., and Europe.  PerceptIn's ultimate goal is to provide affordable and reliable autonomous driving technologies that can allow PerceptIn's operators to generate profits, and subsequently grow the business. In 2017 and 2018, PerceptIn conducted over 10 pilot projects globally to understand the micromobility market, the customer needs, as well as the cost structure of this business.

One of PerceptIn's pilot projects took place in 2017 at ZTE's industrial park in Shenzhen China. ZTE is a leading Chinese telecom company with an enormous campus in Shenzhen, and the campus is filled with over 30,000 workers with tremendous intra-campus transportation needs. In this pilot project, PerceptIn's pods transferred ZTE's workers across the campus. Each PerceptIn's pod packs four high-definition cameras, four mid-range radar sets, and 10 ultrasound sensors, as well as GPS and sensors for wheel odometry [4]. From these pilot projects, PerceptIn collected sufficient operation data and customer feedbacks for internal business analysis.

Based on internal business analysis, if PerceptIn can provide low-speed autonomous vehicles under $70,000 per unit, PerceptIn could generate a reasonable return-on-investment (ROI) for PerceptIn's customers, the autonomous vehicle operators. Thus, in 2018 PerceptIn set the goal to develop autonomous vehicles that can be sold at a price tag $70,000, which is five to ten times lower than what is commonly believed to be possible for commercial autonomous vehicles. However, the $70,000 price tag also imposes very strict and challenging constraints on the design of low-speed autonomous vehicles. In detail, we have to break down the $70,000 into Non-recurring engineering (NRE) cost such as research and development, recurring costs such as the cost of the chassis, the cost of drive-by-wire conversion (meaning to convert a traditional vehicle into one that can be controlled by computers), the cost of sensors, the cost of integration, the cost of customer service, and finally the cost of the computing system [5].

**Situation**: in June 2017, based on the initial feedbacks from the ZTE case study, PerceptIn conducted a study on autonomous driving computing systems [6], PerceptIn concluded that computing is the bottleneck for the commercial deployment of autonomous vehicles, and PerceptIn needed a computing system that is reliable, affordable, high-performance, and energy efficient. Most importantly, we needed a solution that is cost effective and has a short time-to-market. PerceptIn faced several options:

1. Optimization of commercial off-the-shelf mobile System-on-Chip (SoC) computing systems: This approach brings several benefits, first, since mobile SoCs have reached economies of scale, it would have been most beneficial for PerceptIn to build its technology stack on affordable, backward-compatible computing systems. Second, PerceptIn's vehicles target micromobility with limited speed, similar to mobile robots, for which mobile SoCs have been demonstrated before. However, an extensive study is required to fully understand mobile SoCs' suitability for autonomous driving, this may delay PerceptIn's product launch by six months.
2. Procurement of specialized autonomous driving computing systems: there were commercial computing platforms specialized for autonomous driving, such as those from NXP, MobilEye, and Nvidia. They are mostly Application-Specific Integrated Circuit (ASIC) based chips that provide high performance at a much higher cost. For instance, the first-generation of Nvidia PX2 system costs over $10,000. Besides the cost issue, these computing systems mostly accelerate only the perception function in autonomous driving, whereas PerceptIn require a system that optimizes the end-to-end performance.

3. Development of proprietary autonomous driving computing systems: developing a proprietary computing system guarantees that PerceptIn have the most suitable system for PerceptIn's customers and for its workloads, but also means that PerceptIn need to invest a significant amount of financial and personnel resources on this project. Also, the investment does not guarantee the success of this project. It is a huge and risky bet for a startup like PerceptIn.

Table 1: comparisons of the three available options

|  | Affordability | Backward Compatibility | Suitability | Risk |
| --- | --- | --- | --- | --- |
| Option 1: mobile SoCs | HIGH | HIGH | UNKNOWN | LOW |
| Option 2: specialized computing systems | LOW | LOW | MEDIUM | LOW |
| Option 3: proprietary computing systems | LOW | HIGH | HIGH | HIGH |

**Decision**: then the decision process started, without quantitative evaluation methods, we had to use judgement calls to evaluate different options [10, 11]. We summarized these options in Table 1, the parameters of evaluation include *Affordability*, *Backward Compatibility*, *Suitability* for autonomous driving computing, and project *Risk*. For all parameters, the higher the better. For *Affordability*, option 1 is the clear winner. For *Backward Compatibility*, both option 1 and option 3 deliver good results. For *Suitability*, option 3 is the clear winner, and option 1 is unknown and requires additional study. For *Risk*, both option 1 and option 2 have low risk.

Using table 1, PerceptIn quickly ruled out option 2 due to its low score across all parameters except *Suitability*, it is clear from available data that option 2 was the worst choice among the three. Then internal debate began within PerceptIn on whether to move forward with option 1 or option 3. Option 1 seemed very attractive, but it was unknown whether it would be suitable for autonomous driving computing tasks. If it was not suitable, then option 1 would be infeasible. To have a clear answer on whether option 1 would be suitable, a six-month study would be necessary. For option 3, it would be an extremely expensive option and with high risk. Even if PerceptIn invested in this project, there would be no guaranteed success.

Since there is an unknown parameter in option 1, a deterministic analysis could not be conducted, and the PerceptIn management team had to use judgment calls not susceptible to analysis in our decision process.

After several rounds of internal debate, the PerceptIn management team decided to compare option 1 and option 3 in their respective worst case scenario. If PerceptIn moved forward with option 1, the worst case that could happen was that six months wasted, but with limited investment. If PerceptIn did find out that option 1 would not be suitable, PerceptIn could still try option 3. If PerceptIn moved forward with option 3, the worst case that could happen was that half of PerceptIn's R&D budget and 12 months wasted. With this analysis, PerceptIn decided to take the safest approach, also an approach that everyone was comfortable with: move forward with option 1 for six months, if that does not work, then try option 3.

**Results**: for six months, PerceptIn focused on option 1 but unfortunately PerceptIn found that mobile SoCs are ill-suited for autonomous driving for three reasons:

1. the compute capability of mobile SoCs is too low for realistic end-to-end autonomous driving workloads. Figure 1 shows the latencies and energy consumptions of three perception tasks— depth estimation, object detection, and localization—on an Intel Coffee Lake CPU, Nvidia GTX 1060 GPU, and Nvidia TX2, which represents today's high-end mobile SoCs. Fig.1a shows that TX2 is much slower than the GPU, leading to a cumulative latency of 844.2 ms for perception alone. Fig.1b shows that TX2 has only marginal, sometimes even worse, energy reduction compared to the GPU due to the long latency.
2. mobile SoCs do not optimize data communication between different computing units, but require redundant data copying coordinated by the power-hungry CPU. For instance, when using DSP to accelerate image processing, the CPU has to explicitly copy images from sensor interface to DSP through the entire memory hierarchy. PerceptIn's measurement shows that this leads to an extra 1 W power overhead and up to 3 ms performance overhead.
3. traditional mobile SoCs design emphasizes compute optimizations, while PerceptIn finds that for autonomous vehicle workloads, sensor processing support in hardware is equally important. For instance, autonomous vehicles require very precise and clean sensor synchronization, which mobile SoCs do not provide.

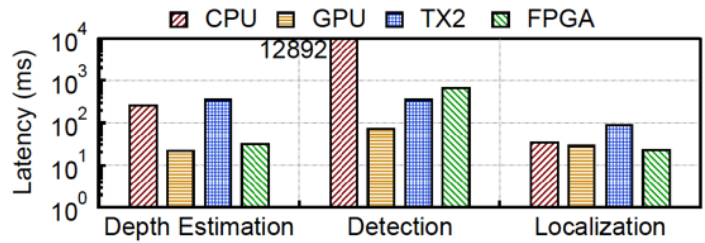

(a) Latency comparison.

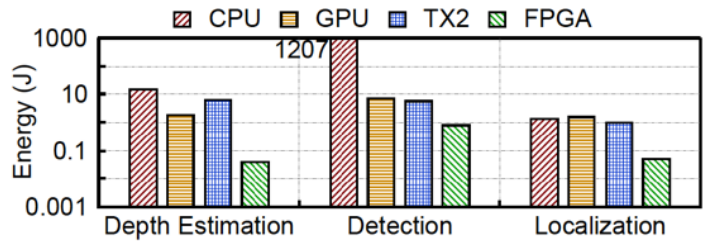

(b) Energy consumption comparison.

Figure 1: performance and energy consumption comparisons.

Starting in early 2018, PerceptIn decided to move forward with option 3 since option 1 proved not to work. PerceptIn thus formed a team to develop the FPGA-based DragonFly computing system [7, 8, 9, 12]. Option 3 was a huge success, today all autonomous vehicles shipped by PerceptIn are empowered by PerceptIn's proprietary DragonFly computing system. However, if PerceptIn had boldly moved forward with option 3 at the beginning, PerceptIn would have been able to ship this great product six months earlier. The main reason of option 3's success was that when using judgement call, we overestimated the technical risk. Of course, during the development process of option 3, we encountered a lot of technical unknowns, and fortunately we found solutions for these technical problems. However, during the planning stage, using judgement call, we overestimated the difficulties of these technical problems, and thus led us to believe that starting with option 1 was a safer approach.

**Retrospective**: PerceptIn shipped its products globally, but delayed by six months because PerceptIn took an R&D detour. At the planning stage, when using judgement call, PerceptIn overestimated the technical difficulties in option 3, and chose to start with an safe path, option 1. In a way, PerceptIn's evaluation was too pessimistic, had PerceptIn taken the hard choice to go with option 3 initially, PerceptIn would have a much higher market share today. By making the decision of going with option 3, PerceptIn would have greatly widened its moat, and enlarging its edge over competitions. In retrospective, the root cause was that the PerceptIn team did not have proper anchoring and adjustment when using judgement call [11], and hence the decision was too pessimistic. The lesson from this project is that, for a technology startup in hypergrowth mode, when its management team identify a critical technical problem, the startup should always choose the best solution over the safest solution. If the team is not comfortable with its decision, outside expert opinion should be enlisted to help provide proper anchoring, this will definitely give the startup a huge advantage over its competitions who chose the easiest solution.

## Refences